\documentclass[12pt]{article}

\usepackage{newtxtext,newtxmath}

\usepackage{jabbrv} 
\usepackage{graphicx}

\usepackage[letterpaper,margin=1in]{geometry}

\renewenvironment{abstract}
	{\quotation}
	{\endquotation}

\date{}

\makeatletter
\renewcommand{\fnum@figure}{\textbf{Figure \thefigure}}
\renewcommand{\fnum@table}{\textbf{Table \thetable}}
\makeatother

\usepackage{scicite}

\usepackage{url}

\def\scititle{
	Chondrite Parent Bodies as Escaped Satellites of Proto-Planetary Embryos
}
\title{\bfseries \boldmath \scititle}

\author{
    Harold F$.$ Levison$^{1\ast\dagger}$,
    Rogerio Deienno$^{1\dagger}$,
    Kevin J$.$ Walsh$^{1\dagger}$,
    Brandon C.~Johnson$^{2,3\dagger}$,\and
    Harold C.~Connolly Jr.$^{4,5\dagger}$,
    Shigeru Wakita$^{2\dagger}$,
    Robert E.~Grimm$^{1\dagger}$\and
    \small$^{1}$Solar System Science and Exploration Division, Southwest Research Institute, Boulder, CO USA.\and
    \small$^{2}$Department of Earth, Atmospheric and Planetary Sciences, Purdue University, West Lafayette, IN, USA.\and
    \small$^{3}$Department of Physics and Astronomy, Purdue University, West Lafayette, IN, USA.\and
    \small$^{4}$Department of Geology, Rowan University, Glassboro, NJ, USA .\and
    \small$^{5}$Department of Earth and Planetary Sciences, American Museum of Natural History, Central Park West, New York, USA.\and
	\small$^\ast$Corresponding author. Email: hal.levison@swri.org\and
	\small$^\dagger$These authors contributed equally to this work.
}

\begin{document} 

\maketitle

\begin{abstract} \bfseries \boldmath
Chondrites are composed of {\color{black}formerly} partially molten material, known as chondrules, surrounded by fine-grained matrix.  They date from the earliest times in Solar System history.  However, their role in the formation of the planets is uncertain because, in part, it is not clear how they were produced. {\color{black}Here we show a robust pathway for forming meteorite-producing asteroids that contain chondrules through embryo–embryo collisions during the late stages of terrestrial planet formation.}  Melted material from these impacts cool into chondrules and mix with unmelted material in embryo-centric disks that formed from the ejecta. This material accretes into numerous asteroid-sized satellites. These objects are later ejected onto heliocentric orbits due to gravitational encounters with other embryos, thereby becoming the parent bodies of chondrites. This mechanism provides a pathway to form chondrites in solar system history at times commensurate with measured chondrule ages, while explaining many of their physical properties.

\medskip\medskip

\noindent Teaser: A model is presented where chondrite parent bodies are produced during collisions between growing planets.

\end{abstract}

\subsection*{{\color{black}Introduction}}

\noindent
Chondrules are expected to play an important role in tracing the events of the early Solar System, and yet it is still unclear whether they were a key precursor of planet formation or an outcome. That this is not yet known is a credit to their confounding physical properties and uncertainty about the evolution of the early solar system (see Review in \cite{Connolly&Jones2016}). Their abundance in meteorites that are linked to a wide range of asteroid types suggest that their formation was widespread both in location and time throughout the extent and lifetime of the gaseous solar nebula. On the other hand, their physical and chemical properties suggest that formation events and their eventual incorporation into parent asteroids (i.e$.$ chondrite parent bodies or CPBs) may have been relatively localized, as we discuss in more detail below.

There has been a veritable cornucopia of possible pathways to form chondrules suggested in the literature, many of which satisfy some of their constraints, such as cooling times, physical sizes and measured ages. Matching all of their constraints in the context of known or estimated solar system evolution is still, seemingly, out of reach (see \cite{Connolly&Jones2016}).  A promising mechanism involves chondrules forming in impacts \cite{Asphaug+2011,Sanders&Scott2012}, especially between planetary embryos \cite{Johnson+2015}. In particular, hydrodynamic simulations find that as much as $\sim\!6$\% of an impactor's mass can be melted and jetted above escape speeds for impacts between embryos \cite{Johnson+2015, Wakita+2021}.  These models successfully reproduce many of the physical aspects of the chondules, themselves. 

The impact jetting model described above can potentially explain the ages of chondrules as well. The putative ages of chondrules peak from around 2 Myr to 4 Myr after Ca-Al-rich inclusion (hereafter CAI) formation in the inner solar system, and up to around 5 Myr in the outer solar system \cite{Kita&Ushikubo2012, Villeneuve+2009}.  However, models of planet formation do not necessarily support the formation of planetesimals in the terrestrial planet region after roughly 2 or 3 Myr years, because they predict that the vast majority of solid material would have already been incorporated into embryos and proto-planets (e.g$.$\cite{Walsh&Levison2019}) or would have simply been lost as they fell into the Sun \cite{Lichtenberg+2021, Morbidelli+2022, Izidoro+2022}, and Jupiter would have likely cut off the supply of pebbles and dust from the outer solar system \cite{Kruijer+2020}.

The impact jetting model solves this problem by having chondrules form during the final assembly of the planets.  Exactly where and when chondrules form according to this model depends on the number, sizes and timing of impacts. The procedure typically used for testing the viability of jetting as a formation mechanism involves a Monte Carlo model of terrestrial planet formation that tracks each impact so that the total expected chondrule mass can be summed over time \cite{Johnson+2015}. This type of model specifically tracks the melted mass of material that is launched above the escape speed of the target body so that it is mixed into the solar nebula. This material cools in a time frame to be consistent with the igneous textures of chondrules. The model assumes that the solar nebula would then be needed to promote the collection of the chondrules into chondrites (through an unspecified process).  Thus, we would expect that the ages of chondrules to be in line with the typical expectation for the lifetime of the solar nebula of 2 -- 5 Myr (e.g$.$ \cite{Weiss+2021} based on paleo-magnetism constraints, estimate the solar nebula dispersed sometime between 1.22 Myr and 3.94 Myr in the inner solar system and between 2.51 Myr and 4.98 Myr in the outer solar system), which is observed.

The story developed in the impact jetting model ends here, more or less, with the potential for chondrules to be ejected out into the wild early days of the solar system where planet formation was actively underway. But many things still need to happen. In particular, they need to be incorporated into sizeable CPBs.  The compositional and isotopic variation observed within a given chondrite implies that its constituent material is genetically related and its bulk composition is solar minus some volatile loss \cite{Bland+2005, Budde+2016}.  This is difficult to understand if chondrules enter heliocentric orbit because chondrule formation events universally produce much less mass than the solid mass in the annulus of the proto-planetary disk in which it occurs. As a result, chondrules will mix, and thus be diluted by, other material from this disk if they are allowed to enter heliocentric orbit without some process to confine them.

An intuitive way to explain numerous constraints would be if chondrules and matrix were sourced at the same time from the same place by the same event. If they could then be corralled in space and time to accrete into distinct asteroid-sized bodies (D$\sim$100km) many of their properties may naturally be explained.  The scenario explored here envisions embryo-embryo collisions during terrestrial planet formation as the “event”.  As for the “corral”, we hypothesize that the CPBs accrete in circum-embryo disks that form from the ejecta of these collisions.  It is  widely accepted that large collisions in the terrestrial planet region are capable of forming circum-planetary disks via models of the Earth-Moon impact (for example \cite{Canup2008}) and Phobos/Deimos formation at Mars \cite{Canup&Salmon2018, Marinova+2011}.  These models show that material in this disk quickly accrete into satellites --- the size of which depends on the surface density of the disk \cite{Canup&Salmon2018}.  These satellites can become the chondritic asteroids we see today, if: 1] they are large enough to survive later Solar System collisional evolution (D$\gtrsim$30km; \cite{Bottke+2005}), but do not differentiate (which occurs only if they grow larger than $\sim 300$km at $\sim 1.5$Myr after CAI formation;  \cite{Lichtenberg+2021, Grimm&McSween1993, Elkins-Tanton+2011}), and 2] they can be dynamically liberated from their embryo-centric orbits during the final throes of planet formation. As we describe in detail below, here we design and perform a series of numerical experiments to explore aspects of this scenario in detail. It is important to note, however, that here we assume that the impact jetting model can adequately explain the characteristics of the chrondrules of interest.  We are not re-litigating this model in what follows. Also, the jetting model assumes that the growing embryos maintained a chondritic upper layer, as argued in Ref.~\cite{Weiss&Elkins-Tanton2013}.

Before we develop our scenario further, we need to discuss the key constraints that guide the scope and parameters of the modeling and data analysis presented here.  In this work we specifically focus on the terrestrial planet region because the models of planet formation are best developed there. The terrestrial and giant planets formed from distinct reservoirs that have different compositions and that these differences survived the planet formation process (for example, see a recent discussion in Ref.~\cite{Morbidelli+2025}) --- showing that there was little mixing between the reservoirs as the planets formed. Indeed, this understanding is what allows models of terrestrial planet formation to only consider material from inside of a few au. Similarly, these models show that compositional gradients are somewhat preserved during this process. (The compositional differences between the Earth and Mars also supports this idea.) Thus, we can employ terrestrial planet formation models to study ordinary chondrites (OCs) meteorites and only consider the properties of the chondrules in OC meteorites in what follows.  We will also briefly discuss the possible implications for other chondrite types. The key constraints or properties of OC's that drive this analysis, and that are utilized to process existing planet formation models, are as follows:

\begin{enumerate}

\item The cooling rates for chondrules are observed to be 10 -- 1000 K/hr \cite{Lofgren&Russell1986, Desch+2012}, which imply that they cooled in a low density environment.  The jetting model in Ref.~\cite{Johnson+2015} can generally explain these rates.

\item Most condrules in OCs formed between $\sim$2 -- 4 Myr after CAIs \cite{Kita&Ushikubo2012, Villeneuve+2009}.
  
\item Chondrules account for 60 -- 80\% mass of ordinary chondrite meteorites  \cite{Desch+2018, Jones2024}.
  
\item Chondrites have essentially solar composition, apart from some volatile depletion. However, in unequilibrated chondrites, the chondrules and matrix have different compositions.  This implies that these components must have remained together in their solar proportions.

\item Around 15\% of chondrules show evidence for multiple heating events \cite{Connolly&Jones2016}.

\end{enumerate}

\subsection*{Forming Embryo-Centric Disks with Chondrules}

The viability of the scenario introduced above depends on: 1] the outcomes of embryo-embryo collisions, 2] their production of jetted melt material, 3] the evolution of bound ejecta forming a circum-embryo disk, 4] the formation of asteroid-sized bodies in the circum-embryo disk, and 5] their eventual liberation onto heliocentric orbits. We take each of these processes in turn. We start with the formation of the disks by interpolating expected outcomes of embryo-embryo impacts as a function of impactor and target masses and the impact properties (speed and angle). We then utilize previous published models of terrestrial planet formation to calculate ranges of expected outcomes.

\medskip
\noindent {\bf Producing melt in an impact and then into the disk:}  A giant impact phase happens in nearly all models of terrestrial planet formation (e.g$.$ \cite{Raymond&Morbidelli2022}) and is essentially mandated by the formation of the Earth-Moon system (see \cite{Canup2013}). Nominally, orbital eccentricities and inclinations are excited in systems containing more than one growing planet because of mutual gravitational interactions.  However, these eccentricities and inclinations are damped while the gaseous solar nebula is present. As a result, impact speeds are very close to the target bodies' escape velocities, and are controlled primarily by the target body's size. The expectation is that the solar nebula was present during the formation of chondrules \cite{Johnson+2016, Weiss+2021}, and therefore we can assume that impact speeds are set by the escape speeds of target bodies.

Melt production increases with impact speed \cite{Wakita+2021}, and models with impact angles larger than $45^\circ$ and speeds larger than 5 km/s produce ejecta that is more than 60\% melt. Thus, we expect a similar fraction could also hold for the ejecta below the target escape speed with impacts faster than 5 km/s.  This material will be available to form embryo-centric disks, as described below.  Impact speeds this high imply targets the size of Mars (whose escape speed is 5.03 km/s).

Terrestrial planet formation simulations show that Mars-size objects are typical targets of large collisions. Indeed, models predict that systems usually go through a stage consisting of  roughly 10-20 Mars-sized embryos that will eventually combine to form the final four planets (\cite{Raymond+2009, Walsh&Levison2019}; however see Ref.~\cite{Johansen+2021} for an exception that does not work in our scenario). 

\begin{figure}
\centering
\includegraphics[width=.85\linewidth]{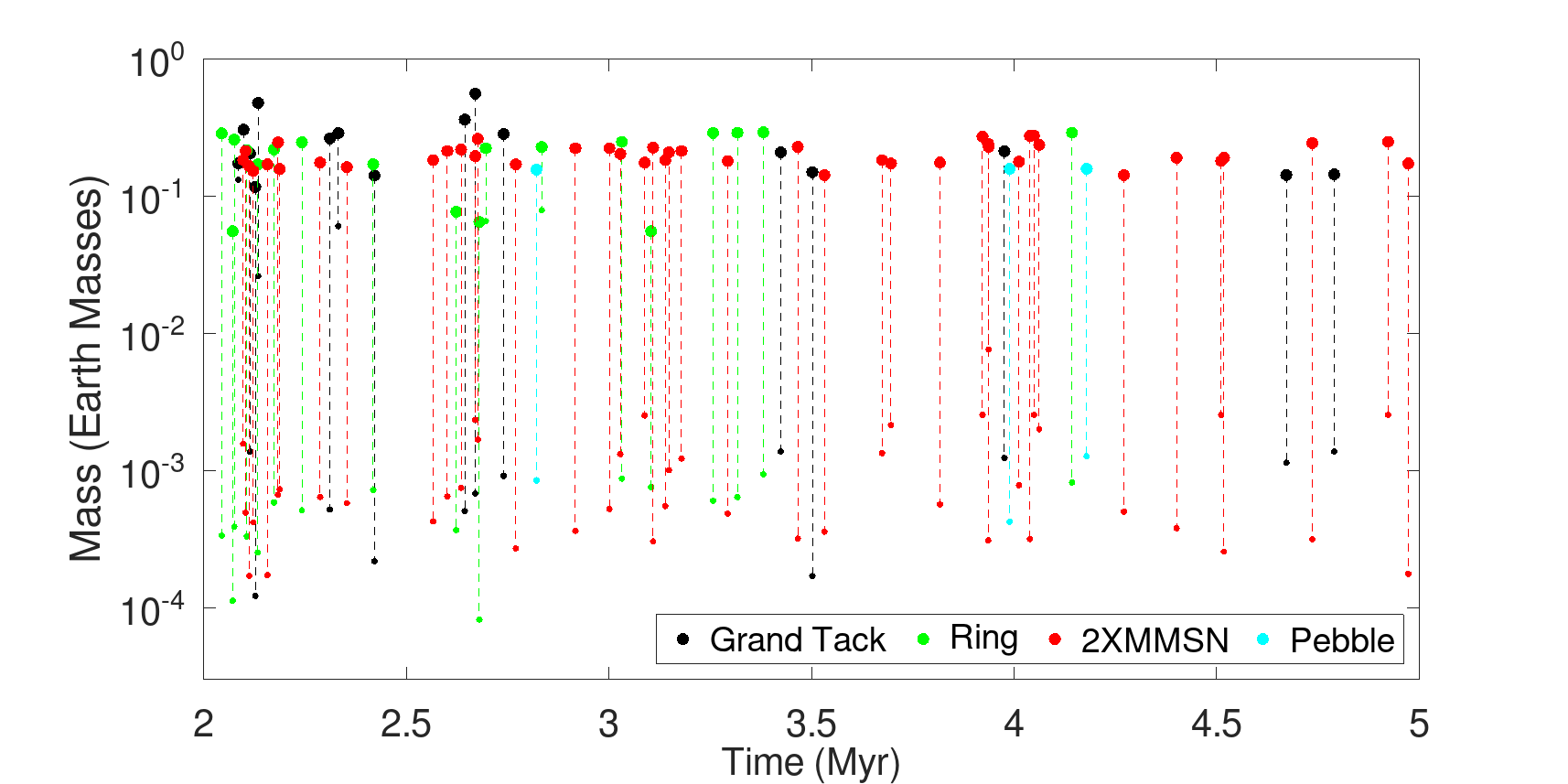}
\caption{\textbf{Embryo impacts with velocities larger than 5 km/s in our representative simulations between 2 and 5 Myr.} The four representative simulations are indicated by the color of the symbols, see the legend. In particular, each impact is represented by two dots. The large and small dots show the mass of the target and impactor, respectively. Each pair is connected by a doted line.}
\label{fig:Impacts}
\end{figure}

In this work we analyze four different scenarios of terrestrial planet formation proposed in the literature. Those scenarios are: $i$] the “Grand Tack” model \cite{Walsh+2011} which accounts for Jupiter's possible transit through the asteroid belt region starting at 2 Myr and lasting roughly 100,000 yr, $ii$] the ``Ring" model (as we call it here) where we follow growth from planetesimals starting from the conditions of the inner ring described in Ref.~\cite{Izidoro+2022}, $iii$] the 2 times minimum mass solar nebula case (2xMMSN hereafter; \cite{Walsh&Levison2019}), and finally $iv$] the pebble accretion model as proposed in Ref.~\cite{Levison+2015}, in which several Moon- to Mars-sized embryos form on their way to became the current observed Mercury, Venus, Earth/Moon, and Mars. All of these models are successful in reproducing our current terrestrial planet system main characteristics, i.e., orbital and mass distributions. We analyze one representative model, which was chosen at random, from each of these scenarios. Figure~\ref{fig:Impacts} shows all the impacts between 2~Myr (the earliest OC chondrules) and 5~Myr (the assumed lifetime of the gas
disk) with velocities greater than 5 km/s (for reasons described below) for each of these simulations.  These are the collisions that can produce chrondrules in the abundance observed in OCs. 

Although each giant impact that occurs in these simulations is shown separately in Figure~\ref{fig:Impacts}, it is important to note that several of these impacts occur on the same target body.  As we describe in more detail below, this result allows for the possibility of creating CPBs that contain chondrules with different impact ages and compositions.

\medskip
\noindent {\bf Circum-embryo disks are likely outcomes of embryo-embryo impacts:} Now that we have demonstrated that impacts with objects larger than Mars were common at the time when chondrules were forming in the early solar system, we turn our attention to whether these impacts will form disks.  Disks are a generic outcome of embryo-embryo collisions as found in models of the Earth-Moon impact \cite{Ida+1997, Canup2008} and a wide range of impacts on Mars \cite{Citron+2015, Canup&Salmon2018}. Significant modeling of the Earth-Moon impact provides a pathway to estimate which impacts lead to circum-embryo disks and how much mass is in those disks. For example, an impactor of $0.1 M_\oplus$ hitting the proto-Earth at an angle near $45^\circ$ with a relative velocity at infinity of no more than 10\% larger than the Earth’s escape speed will place roughly two lunar masses into a disk \cite{Canup2008}. For smaller impactors, studies found that impactors as small as $\sim 10^{-3}$ the target mass can lead to disks that will have a total mass approaching $10^{-4}$ the total mass of the system \cite{Canup2008, Canup&Salmon2018}. 

\begin{figure}
\centering
\includegraphics[width=.8\linewidth]{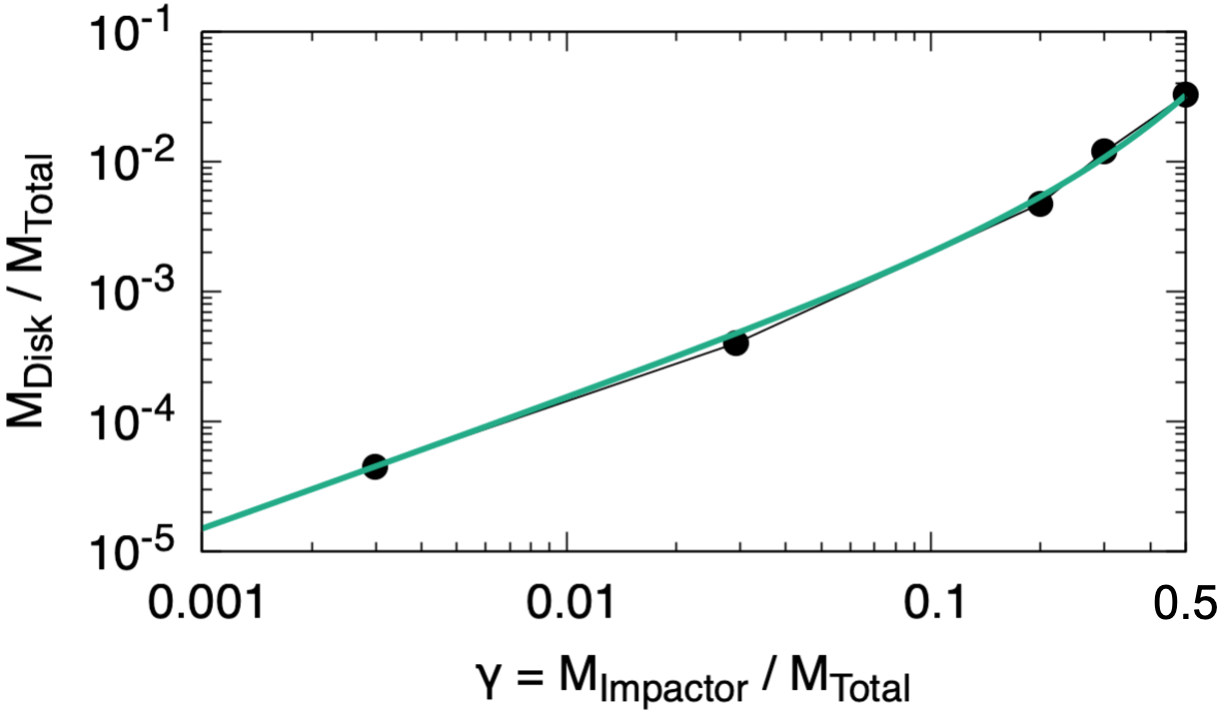}
\caption{\textbf{Impact generated disk mass as a function of $\gamma$.}  The black dots show the results from published SPH simulations \cite{Canup2008, Canup&Salmon2018}. The green curve is a fit to data points as described in the text.}
\label{fig:Diskmass}
\end{figure}

The mass of the disk depends primarily on the fraction of the total mass contained in the impactor ($\gamma \equiv$ M$_\mathrm{impactor}$/M$_\mathrm{total}$). Simulations of large impacts into the Earth and Mars  in Refs.~\cite{Canup2008} and \cite{Canup&Salmon2018} find that M$_\mathrm{disk}$/M$_\mathrm{total}$ increases from 10$^{-5}$ to $\sim$3$\times$10$^{-2}$ as $\gamma$ increases from 0.001 to 0.5 (the impactor is the same mass as the target; see Figure~\ref{fig:Diskmass}).  We employ these results in our analysis of our new chondrite scenario, which is developed below.  To accomplish this, we performed a fit to the data points in Figure~\ref{fig:Diskmass} (the green curve),
\begin{equation}
  {\rm M}_{\rm Disk} / {\rm M}_{\rm Total}  = 0.015\gamma e^{2.92\gamma},
\end{equation}
that allows us to determine the disk mass for any collision that occurs in our planet formation simulations.  

Meanwhile, these SPH simulations (Ref.~\cite{Canup&Salmon2018}, for example) show that the disks typically have steep surface density profiles --- $\Sigma(r) = \Sigma_{\rm 0}r^{\rm -x}$ with $x \sim 4.5$.  In what follows, we assume that the disk that forms from impacts follows this surface density profile.

\medskip
\noindent {\bf Producing melt in an impact and then into the disk:}  Models of the jetting origins for chondrules have previously been analyzed for the amount and properties of jetted material that is ejected {\it above} escape speed \cite{Johnson+2015, Wakita+2021}. These have been done for different material properties, impact speeds and angles. To evaluate our new scenario, we are required to know the amount and properties of material that is below escape speed, but above circular velocity at the surface; i.e$.$ the material that can reach orbit.   Ref$.$~\cite{Johnson+2015} has shown that melt cools to form chondrules, and so an estimate of melt fraction in a simulation provides an measure of percentage of the generated ejecta that is composed of chondrules. 

To quantify the amount of melt/chondrules (we refer to this as `melt' for the remainder of this discussion) that remains in orbit about a target, we expanded the jetting simulations in Ref.~\cite{Wakita+2021}, using procedures described Materials and Methods. We find that melt production increases with impact speeds (Figure~\ref{fig:Meltfraction}), and the amount of chondrules that is deposited in circum-embryo disks becomes non-zero above 3 km/s. Our models also show a strong dependence on target porosity, since the porous case produces more melt \cite{Melosh1989}.   In what follows, we require the melt fraction to be more than 60\% by mass in order to match the observed chondrule mass fraction found in meteorites, which implies impacts with speeds larger than 5 km/s. 

Note, that the melt fraction is derived separately from disk mass, where the former is from a series of simulations specific to modeling jetting (Figure \ref{fig:Meltfraction}), and the latter is from a large collection of previously published works focused on satellite formation at different worlds (Figure \ref{fig:Diskmass}). In absolute terms the predicted amount of bound material from jetting modeling is similar to the predicted disk mass from satellite formation modeling, giving confidence to this approach that relies on such a wide variety of modeling techniques. For this study, the disk mass for any large impact is derived from the fit in Eq.~[1] (shown in Figure \ref{fig:Diskmass}), and the fraction of the disk that is melt-derived chondrules comes from the jetting modeling outputs in Figure \ref{fig:Meltfraction}. 

\medskip
\noindent {\bf Producing chondrule-rich disks during terrestrial planet formation:}  In this section we combine the pieces of the puzzle developed above to determine the characteristics of the disks produced in collisions, and the total mass of material in these disks. Recall that the observed ages of chondrules in ordinary chondrite meteorites imply that the disks must form between 2 and 4 Myr after CAIs.  Also, these same meteorites are typically composed of more than 60\% chondrules by mass, which indicates that at least this fraction of the resulting disks must be constructed of material that had been melted. For this calculation we  assume 25\% porosity for the target in the calculation of melt fractions. It is mandatory for matching the physical properties of the chondrules that the target body is not yet fully differentiated \cite{Bland+2005} and, at least, has a thin primitive undifferentiated outer shell \cite{Weiss&Elkins-Tanton2013}, and so moderate porosity at its surface is reasonable. It should be noted, however, that this is the most optimistic outcome from the jetting modeling.

The planet formation simulations described above are post-processed to tabulate all collisions above V$_{imp}\geq$ 2 km/s and their coordinates and velocity vectors are recorded. The four terrestrial planet simulations were originally performed for different lengths of time.  As a result we only consider impacts that occurred before 5~Myr to ensure the results are comparable.  The functions shown in Figures \ref{fig:Diskmass} (Eq.[1]) and \ref{fig:Meltfraction} are then used to convert the recorded collisions into mass of the circum-embryo disks created and what fraction of that mass formed from melt (which we now refer to `chondrules'). 

The accumulated amount of mass in circum-embryo disks as a function of time is tabulated for disks that contain different chondrule fractions.  This was done for two separate starting times: the beginning of the simulation (dashed curves in Figure~\ref{fig:MassTime}) and at 0.8~Myr (solid curves in the figure).  The later was chosen because before that time, any satellite larger than 30~km would differentiate \cite{Grimm&McSween1993,Lichtenberg+2021} and thus would not be contributing to the chondrule population. In what follows we only consider impacts that occurred after 0.8~Myr.  The total mass in circum-embryo disks that contain more than 60\% chondrules (solid yellow) varies largely among the four models. Not all circum-embryo disks created reach the 60\% criteria (solid blue in Figure~\ref{fig:MassTime}) and those with lower chondrule fractions would produce asteroids that do not match typical constraints from OC-chondrites.  Therefore, it is important to evaluate whether the total accumulated mass in `high-chondrule' (more than 60\% condrules by mass) circum-embryo disks is substantially more than the total mass of circum-embryo disks with chondrule fractions that would not match constraints (`low-chondrule' -- less than 60\% by mass). This only happens to the models presented in panels (A) and (B) of Figure \ref{fig:MassTime}, i.e., the Grand Tack \cite{Walsh+2011} and the Ring \cite{Izidoro+2022} models. The 2xMMSN and Pebble Accretion models (panels C and D in Figure \ref{fig:MassTime}) fail this basic constraint and thus are not viable.   

Rejecting models that produce a large fraction of low-chondrule disks is a simplification, albeit a conservative one.  Figure~\ref{fig:MassTime} shows that most of these disks form early enough in the history of the solar system that heating from short-lived radionuclides, i.e.~$^{26}$Al, is likely to still be important \cite{Lichtenberg+2021, Grimm&McSween1993, Elkins-Tanton+2011}.  For example, we have applied our model of satellite formation developed below to a productive low-chondrule disk that formed in the Grand Tack simulation near 1Myr (identified by the step seen in the blue and black solid curves in Figure~\ref{fig:MassTime}a).  We find that $>99.9\%$ of the disk material is accreted into satellites that have diameters larger 100 km, and thus should differentiate according to Ref.~\cite{Grimm&McSween1993,Lichtenberg+2021}.  So while these disks will not be contributing to the population of chondrites,  they could be responsible for some of the parent bodies the iron meteorites.  Thus, terrestrial planet formation models that produce a large amount of low-chondrule disks might still be viable.  However, for simplicity sake, we ignore them here.

The above discussion might tempt us to conclude that both the Grand Tack and the Ring models are the only ones tested that are compatible with our proposition. We caution against drawing any conclusion about which planet formation scenario is viable because we only studied one realization from each.  The calculations presented here are complicated and time consuming.  Because the purpose of this work is to show that our CPB formation mechanism is plausible, we only performed enough simulations to demonstrate that fact.  Having said this, there are dynamical characteristics that our successful models possess that are missing in the unsuccessful ones, which are likely to be good indicators of viability.  Only those models with many large collisions among Moon- to Mars-sized embryos are able to generate significantly more high-chondrule circum-embryo disks than low-chondrule circum-embryo disks.  All our models form a first generation of terrestrial planets before 2--3 Myr that consist of a series of Moon- to Mars-sized embryos.  As the surface density of the gas decreases, these systems eventually become excited enough for embryo-embryo impacts to occur ---  leading to the final assembly of the planets.  Our successful models either first produced more compact first-generation planetary systems or had first generation planets pushed together by the dynamical evolution of the giant planets. The more compact orbital configurations allow for more violent dynamical excitation during the final assembly of the planets, and thus higher velocity impacts and more melt being formed during collisions (Figure \ref{fig:Meltfraction}).

Despite the caveats outlined in the last two paragraphs, we will only continue to develop the Ring and Grant Tack models because they pass this first important test --- they succeed in generating enough high-melt circum-embryo disks to dwarf those with low-melt.  There are two important takeaways from Figure~\ref{fig:MassTime}.  First, the majority of chondrules in these simulation form after 2Myr.  In particular, the GT simulation creates 97\% of its chondrules after this time, while the Ring simulation produces 85\%.  Thus, these models are consistent with the observed formation times of these objects.  Second, the total cumulated mass in high-melt material generated by these models is about an order of magnitude larger than the total mass of the current asteroid belt (solid green horizontal line in Figure~\ref{fig:MassTime}).  A discussion on the overall efficiency of our scenario can be found in the Discussion section below.

\subsection*{Forming asteroid-sized objects in circum-embryo disks}

The circum-embryo disks that have the appropriate fraction of melted material also need to be capable of building enough CPBs in the correct size range and at the correct times to explain the observed OCs.  We restrict the following analysis to disks that formed between 2 and 5 Myr to be consistent with the observed chondrule ages. Bodies that are larger than $\sim$ 30 km are needed to survive the collisional evolution of the Solar System \cite{Bottke+2005}.  They also cannot differentiate.  Differentiation is not expected to be an issue because the impacts treated here occur later than 2 Myr after CAI formation, i.e., late enough that $^{26}$Al has largely decayed (cf.~\cite{Lichtenberg+2021, Grimm&McSween1993, Elkins-Tanton+2011}, we address this issue again below). Therefore, we do not need to employ an upper limit to the size of our CPBs. The number and sizes of the satellites that form in a disk is a function of several disk parameters; only one of which is the disk mass. Modeling of this complex process has only been done in detail for a few notable cases; e.g. the inner Saturnian satellites, the Moon, and the Martian satellites. 

We construct satellite systems in the circum-embryo disks above using the techniques described in Materials and Methods, and are based, in part, on previous models in Ref.~\cite{Crida&Charnoz2012}, hereafter CC12.  We find that these calculations reach an endpoint where a system of satellites stretches from the Roche limit ($R_r$) at $\sim$2.5 planetary radii to the largest the outer limit set in the simulation at $10 R_r$.   This corresponds to a range in normalized distance, $\Delta \equiv (R/R_r - 1$), from 0 to 9.  The sizes of the satellites varies as a function of distance from the embryos where the largest objects are found near the middle of the disk, for reasons discussed in Materials and Methods. The smallest satellites in a system are usually smaller than the 30 km size cut-off.  These objects are not included in the analysis below.  {\color{black}It is important to note that the CC12 models were developed for satellite growth and accretion in a system without gas (Saturn’s moons), whereas the systems we model here could be affected by the remaining solar nebula. Modeling this affect is beyond the scope of this work. However as we describe in Materials and Methods, the density of the disk solids is orders of magnitude larger than the surrounding nebula and so the presence gas in unlikely to affect our results}.

As discussed previously, the Grand Tack and Ring models are the only ones of the four that satisfy the melt fraction criterion, and so are studied here. Our Grand Tack simulation (Figure~\ref{fig:CC12}A) produced a total of 16 individual circum-embryo disks and cumulatively generated a total of 462 distinct high-melt objects (i.e$.$ CPBs) larger than 30 km, totaling 34 $M_{\rm AB}$).  Note that many of our CPBs are larger than the largest S-type asteroid, and so we assume that subsequent collisions break these objects apart.   Similarly, our Ring simulation (Figure~\ref{fig:CC12}B) produced a total mass of about 9 $M_{\rm AB}$ of satellites in the form of 512 distinct objects from 18 distinct circum-embryo disks. It should be noted that the effectiveness of our chondrite formation scenario is very sensitive to exactly how the terrestrial planets formed, which is uncertain.  We will address this issue again in the Discussion section.

\subsection*{Satellite liberation to heliocentric orbit}

The CPBs that we know of today are not found in orbit about the planets, of course, but in heliocentric orbit.  In this section we model the dynamical evolution of the satellites constructed in the last section during the final assembly of the terrestrial planets.  As the system continues to evolve, planetary embryos will suffer close encounters with one another, potentially leading to existing satellites being gravitationally stripped away from their parent embryos and entering heliocentric orbits. Here we estimate the amount of chondritic mass that arrives on heliocentric orbits by this process using the procedure described in Materials and Methods. Following previous arguments, we performed these calculations on the two terrestrial planet simulations shown in Figure~\ref{fig:CC12}:  the Grand Tack, and Ring.  

The temporal evolution of the total mass of CPBs is presented in Figure~\ref{fig:Strip}.  The black curves show the variation of the instantaneous mass in chondritic satellites (i.e$.$ gravitationally bound to their parent embryo) with time.  The total mass increases when each disk is formed and decreases as satellites are stripped away or collide with their parent embryos.  Between 60\% and 95\% of our satellites by mass evolve onto heliocentric orbits.  The vast remainder collide with their parent embryo.  The Grand Tack, and Ring simulations produce 32.7 and 8.3 M$_{\rm AB}$ of CPBs in heliocentric orbits, respectively.  

The question naturally arises as to whether this process produces enough chrondritic material to explain the S-complex that we observe.  Roughly $1/4$ of main belt asteroids are broadly S-complex taxonomic types \cite{Mothe-Diniz+2003}, implying that we need approximately 0.2 -- 0.3 M$_{\rm AB}$  of OC material delivered to the asteroid belt. The two terrestrial planet models discussed here produce more than that.  However, in order to produce a small Mars \cite{Morbidelli+2012} these models are designed so that embryo-embryo impacts occur closer to the Sun than the asteroid belt, and so our chondrites need to be implanted farther from the Sun than they formed.  Current estimates suggest that this implantation process is inefficient, where only about $\mathcal{O}(1\%)$ might be implanted \cite{Raymond&Izidoro2017, Izidoro+2024}.  If true, only the Grand Tack simulation can explain the entire OC population.  We return to this issue in the Discussion section below.

\subsection*{Discussion}

We suggest that the CPBs currently in the asteroid belt formed during the late stages of planet formation as a result of embryo-embryo collisions.  The CPBs themselves accreted in impact-generated circum-embryo disks that consists almost entirely of material from the impact, thereby naturally resulting in objects that are nearly chemically homogeneous.  The timeline of our scenario is:

\begin{enumerate}

\item {\it Time 0:} CAI formation

\item {\it Time $\approx$1Myr:} The terrestrial planet region becomes isolated from the outer disk \cite{Levison+2015, Kruijer+2017, Kruijer+2020, Izidoro+2022, Morbidelli+2022}

\item {\it Time $\approx$2Myr:} Mars mass embryos appear and the original planetesimals are largely accreted \cite{Dauphas&Pourmand2011, Walsh&Levison2019}

\item {\it Time 2--5Myr:} Solar nebula dispersal \cite{Krot+2005, Pape+2019, Weiss+2021}.  The terrestrial planets become dynamically excited as the surface density of the gas disk decreases resulting in higher speed encounters and impacts.  The impact generation of chondritic, impact-generated satellite systems form and are scattered onto heliocentric orbits.

\item {\it Time $\sim 4 <$ T $<$ 100Myr:} Final terrestrial planets accrete.  Close encounters between the growing planets stop and so does CPB formation.

\item {\it Time 100Myr--today:} After the giant planet instability and implantation into the main belt \cite{Izidoro+2024} there are no further significant dynamical upheavals in the solar system and so the asteroid belt mass is roughly static \cite{Deienno+2024}. 

\end{enumerate}

The scenario and calculations described in this work find that a significant portion of the chondrules estimated to populate the ordinary chondrite parent bodies could have been produced through the described mechanism.  In particular, we find that this mechanism can produce roughly 0.33 M$_{\rm AB}$ of main belt OC asteroids, which is enough to explain the observed population (which consists of 0.2 -- 0.3 M$_{\rm AB}$).  However, only one of the four terrestrial planet models that we studied produced that much material, while creating chondrites that are more than 60\% chondrules by volume.  It is important to note that there are significant uncertainties and approximations in the quantitative models we employ.  Principal among those being that we do not yet understand how terrestrial planets form.  We also have applied models in regimes in which they were not designed (i.e$.$ disk formation),  used processes that did not represent all the important dynamics (i.e$.$ satellite formation), and independently study processes that are actually occurring at the same time.  Given the uncertainties, this work should be viewed as presenting a possible pathway to form OC asteroids rather than a quantitative model.  

This model does make numerous significant and important predictions, and also has numerous caveats or is challenged by some observables or known constraints:

To zeroth order, this model predicts that most chondrules in one meteorite should have the same age. The disks form rapidly around an embryo and from material that would have all been melted at nearly the same instant.   The chondrules within one meteorite should share similar sizes, cooling times, and other basic physical properties owing to their formation from the same materials at the same times in the same ambient environment (i.e., circum-embryo disks).  

However, there are instances in our terrestrial planet simulations (Figure~\ref{fig:Impacts}) where the same embryo is involved in multiple disk-forming impacts over a short period of time.  Although we do not study the evolution of such systems, we can imagine that the resulting chondites will show some internal variation in ages, which may be consistent with the Pb \cite{Bollard+2017} and Al \cite{Siron+2022} isotopic ages of chondrules.  This could also explain why some chondrules show evidence for multiple heating events. In addition, multiple impacts could explain compositional differences in the chondrules in some CPBs \cite{Marrocchi+2024}, for example) because the impactors are different and material from the impactor could dominate the material in the disk \cite{Canup2008}. This scenario predicts, however, we should observe a small number of discrete age/composition types in any chondrite.  Allowing for multiple generations of chondrules within a CPB would relax our assumption that only impacts that produce more than 60\% melt are viable, because the melt fraction in the outer layers of an embryo would build up as the number of collisions accumulate.  As a result, lower velocity collisions, which are ignored in our calculations, could become important thereby increasing the efficiency of the entire process.

The results presented above assume that large satellites ($\gtrsim 300$ km diameter) remain undifferentiated if they formed after 2 Myr because the $^{26}$Al would be sufficiently depleted. Our models fail if $^{26}$Al heating is important well after 2 Myr because nearly all the satellites produced would be differentiated. The Grand Tack model produces the vast majority all of its chondules between 2 and 2.4 Myr after CAI, and none of the models are viable if $^{26}$Al heating last beyond 3 Myr. Isotopic measurements (e.g.~\cite{Kruijer+2020, Lichtenberg+2023}) and thermal modeling (e.g.~\cite{Grimm&McSween1993, Lichtenberg+2023}) allow a range 1.5 to 4 Myr after CAIs for formation of chondrite parent bodies. Given this range of uncertainty, we require (within the limited sample of models here) accretion times around 2--2.5 Myr after CAI. However, note the above discussion of the large uncertainties in our models, and recall that it is possible that these objects might only partially differentiate.

{\color{black}All models of chondrite formation suffer from the so-called “CAI storage problem,” in which a substantial population of CAIs must be stored in the solar nebula for up to 5 Myr and then become incorporated into the CPBs \cite{Wood1996, Ciesla2007}. A similar problem exists for presolar silicate grains, which are found in CPBs but are much older than the chondrules or matrix \cite{Zinner2014}. While addressing this issue is beyond the scope of this paper, we note that these contaminants must be accreted by the growing CPBs. Our scenario has a distinct advantage because material destined to become CPBs will have a substantially larger surface-density-to-mass ratio while it is in the proto-satellite disk than it will as an asteroid. During that time, CAIs and presolar grains can accrete much more effectively. In particular, we find in our satellite formation simulations (cf.~Figure~\ref{fig:CC12}) that the portion of the circum-embryo disk that lies within the Roche limit has an average surface area that is 1200 times larger than the total surface area of the satellites that eventually form. Thus, CAIs and presolar grains can accrete much more effectively.  In addition, these disks are relatively long-lived. We find that they have a median half-life (i.e., the time for the disk to lose 50\% of its mass) of 30,000 years, or a median lifetime of 100,000 years to lose 90\% of their mass.} 

An important implication of the scenario presented here is that chondrules are the result of planet formation --- not the other way around.  Given the measure and modeled short timescales for the growth of Moon- to Mars-sized embryos (e.g.~\cite{Walsh&Levison2019}), depletion of planetesimals and frustrated flow of dust from the outer solar system (e.g.~\cite{Izidoro+2022}), the 2 Myr age of chondrules (e.g.~\cite{Pape+2019}) would be hard to satisfy otherwise. While some outer solar system meteorites do not have chondrules (CI; \cite{Kawasaki+2022, Cashion+2025}), and therefore formed via other processes, this model suggests that nearly all inner solar system chondrites would have formed by this process. In particular, our process likely also applies to carbonaceous chondrites thought to form in the outer solar system \cite{Kruijer+2017}. Carbonaceous chondrites have lower chondrule abundances than ordinary chondrites, which may result from ice reducing chondrule formation efficiency \cite{Cashion+2022}.  Of particular note are the CB and CH chondrites, which show evidence of being formed in impacts \cite{krot+2005b,krot+2021}.

Finally, this means, or could mean, that some of our important chondrite groups look like planets not because they formed them, but the other way around. For example, Enstatite chondrites could have formed in a disk around the proto-Earth, as a direct result of its formation.

\subsection*{Materials and Methods}

{\bf Melt Fraction Calculations:} To quantify the melt fraction of material that enters orbit about a target, we expanded the jetting simulations in Ref.~\cite{Wakita+2021}.  In particular, we use \textsc{iSALE-3D} shock physics code \cite{Elbeshausen+2009, Elbeshausen&Wunnemann2011}. We simulate oblique impacts between a spherical dunite impactor onto a flat dunite target. Note that dunite has been used in chondrule formation studies (e.g$.$ \cite{Johnson+2015}) and similar to ordinary chondrites \cite{Svetsov&Shuvalov2015}. We consider the impact velocities of 2.5, 4, 6, and 7 km/s, in addition to previous work (i.e., 2, 3, and 5 km/s). While previous work focused on non-porous material, we also consider the porosities of 10 and 25\% with varying impact velocities from 2 to 7 km/s. We assume a fixed impact angle of $45^\circ$. Using Lagrangian tracer particles, we track ejecta’s location, velocity, and temperature. When ejecta’s post-shock temperature exceeds the solidus of dunite (i.e., 1373 K), we regard them as melt.  In these new calculations, we specifically tracked the ejecta (both melted and unmelted) with velocity between target's circular velocity and its escape speed.  We employ the ratio of melted to total material in the calculations decribed in the main text.

{\bf CPB Formation Calculations:}  Here we describe our methods for constructing asteroid-sized satellites in our chondrule-rich circum-embryo disks. As discussed in the main text, the expected surface density profile of post-impact circum-embryo disks is quite steep ($\Sigma(r) =\Sigma_{\rm 0}r^{\rm -4.5}$; \cite{Canup&Salmon2018}), with a substantial amount of mass straddling the embryo's Roche Limit at $R_r$.  The material beyond $R_r$ will directly accrete to form satellites.  The material inside this location will viscously spread and slowly form satellites immediately beyond $R_r$ that then migrate outward due to gravitational interactions with the disk  (see Refs.~\cite{Charnoz+2010, Charnoz+2011, Crida&Charnoz2012} for an example for Saturn's rings).   Therefore we construct our satellite systems in two steps.  First, in order to account for the mass originally outside of $R_r$, we  put down a series of satellites beyond the Roche Limit by assuming that each body has reached the so-called isolation mass ($M_{\rm iso}$; \cite{Kokubo&Ida2002}), which is the size that an isolated body would grow if it could accrete all the material in its feeding zone.  The isolation mass is  a function of the distance from the embryo and the local surface density of the disk.  We build our system from the inside out.  The first satellite is placed at the center of an annulus ($R_c$) whose inner edge ($R_i$) is at $R_r$ and whose outer edge ($R_o$) is set so that the mass in the annulus is $M_{iso}(R_c)$.  We then repeat the process after setting $R_i$ of the next satellite to $R_o$ of the last one constructed.  We continue this process until we reach 10 embryo radii, which is consistent with the outer edge of disks produced in the SPH simulations \cite{Canup&Salmon2018}.

We then evolve this system as additional satellites form at $R_r$ as material originally inside the Roche limit spreads outward due to the viscous evolution.  In particular, we employ the model in CC12. In this model, a disk viscosity based on self-gravity and mutual collisions \cite{Daisaka+2001} leads to disk-spreading and thus material crossing the Roche limit.  The flux depends on the size ratio of the disk to the primary body. Once beyond the Roche limit, the material accretes to form satellites that move away from the disk due to tides and resonant interactions. Very large disks, like Earth-Moon or Pluto-Charon, have high fluxes and result in a single large satellite. Smaller disks spread more slowly and can produce several discrete satellites. CC12 provide a prescription for calculating the time evolution of the disk and satellite production, which is applied here to each impact discussed above.  However, in addition to applying these procedures to the satellites created at $R_r$, we apply it to all satellites including those that originally formed outside using the procedure in the last paragraph. 

There is an endpoint that is reached where a system of satellites stretches from $R_r$ to $10 R_r$ or $\Delta = 9$.  The sizes of the satellites varies as a function of distance from the embryos where the largest objects are found near the middle of the disk.  The satellites in the inner part of the disk are formed by the CC12 mechanism and are small because of the mass of the inner disk is decreasing with time, while $M_{\rm iso}$ is small near the outer edge of the disk because of the steep surface density distribution that we employ.  

It is important to note that we are ignoring the effects of nebular gas on the dynamical evolution of the development of the disk and the subsequent accretion of the satellites.  While we acknowledge that this is a simplification and that it should be studied in detail in the future, we believe that ignoring the gas is justified because the bulk density of the solids is much larger than that of the nebular gas and so the solids will drag the gas around rather than the other way around \cite{Weidenschilling1980,Nakagawa+1986,Taki+2016}.  The MMSN is estimated to have an initial mid-plane gas density at 1 au of $1.4 \times 10^{-9} \rm{g}/\rm{cm}^3$ \cite{Hayashi1981}, which puts the gas density at the oldest chrondule ages at $7 \times 10^{-10} \rm{g}/\rm{cm}^3$ for a disk with a exponential lifetime of 3 Myr \cite{Haisch+2001}.  {\color{black}Perhaps the simplest, and most conservative, method of calculating the bulk density of solids is to assume that our thin, centrally concentrated (i.e. recall the surface density $\sim r^{-4.5}$) disk is uniformly spread over a sphere with a radius the same as that of the disk (set to  $10\,R_r$ above).  This calculation grossly underestimates the bulk density because the mass is actually stuffed into a narrow plane and much closer to the embryo.  Even with this extreme approach, we} find bulk densities between $3 \times 10^{-8} \rm{g}/\rm{cm}^3$ and $8 \times 10^{-5} \rm{g}/\rm{cm}^3$, with a mean value of $6 \times 10^{-6} \rm{g}/\rm{cm}^3$ --- orders of magnitude larger than the gas density. 

{\bf Satellite Stripping:}  The goal of these calculations is to estimate the amount of chrondritic material that enters heliocentric orbit from circum-embryo satellite systems via close encounters between embryos.  Unfortunately, although all embryo-embryo collisions were recorded during the four terrestrial planet simulations that we employ as our test cases, close encounters were not.  As a result, we are required to reconstruct the evolution of the system after each disk-forming collision.  This is accomplished in two steps.  

First, for each collision that was deemed to create a disk by the process described above the state of the entire system was used to estimate the eventual liberation of its satellite system. For the timestep immediately following the collision, the state of all embryos (i.e$.$ objects roughly larger than the moon) in the system were extracted from the simulation. The system is then integrated for 10 Myr with a version of \textsc{SyMBA} \cite{Duncan+1998} that tracks close encounters that are closer than 20 times the radius of the parent embryo. This generates a list of close encounters between the parent embryo and other embryos in the system. Note that the evolution of these systems will be different from that of the original system because the dynamics are chaotic and these calculations lack the effects of gas and dynamical friction (which should be small at these late times).  The lack of the later two effects implies that the parent embryo will suffer fewer and faster (i.e$.$ less damaging) encounters and so these new simulations are conservative.

The satellite systems themselves are then integrated through each close encounter.  In particular, the orbit of each satellite is followed in a system that consists of the Sun, the parent embryo, and the interloper.  The satellites themselves do not gravitationally interact.  The orbits of the satellites are assumed to remain unperturbed between encounters.


\subsection*{Supplementary materials}
\noindent None


\clearpage 

%

\newcommand\aj{AJ}
\newcommand\araa{ARA\&A}
\newcommand\apj{ApJ}
\newcommand\apjl{ApJL}     
\newcommand\apjs{ApJS}
\newcommand\ao{ApOpt}
\newcommand\apss{Ap\&SS}
\newcommand\aap{A\&A}
\newcommand\aapr{A\&A~Rv}
\newcommand\aaps{A\&AS}
\newcommand\azh{AZh}
\newcommand\baas{BAAS}
\newcommand\icarus{Icarus}
\newcommand\jaavso{JAAVSO}  
\newcommand\jrasc{JRASC}
\newcommand\memras{MmRAS}
\newcommand\mnras{MNRAS}
\newcommand\pra{PhRvA}
\newcommand\prb{PhRvB}
\newcommand\prc{PhRvC}
\newcommand\prd{PhRvD}
\newcommand\pre{PhRvE}
\newcommand\prl{PhRvL}
\newcommand\pasp{PASP}
\newcommand\pasj{PASJ}
\newcommand\qjras{QJRAS}
\newcommand\skytel{S\&T}
\newcommand\solphys{SoPh}
\newcommand\sovast{Soviet~Ast.}
\newcommand\ssr{SSRv}
\newcommand\zap{ZA}
\newcommand\nat{Nature}
\newcommand\iaucirc{IAUC}
\newcommand\aplett{Astrophys.~Lett.}
\newcommand\apspr{Astrophys.~Space~Phys.~Res.}
\newcommand\bain{BAN}
\newcommand\fcp{FCPh}
\newcommand\gca{GeoCoA}
\newcommand\grl{Geophys.~Res.~Lett.}
\newcommand\jcp{JChPh}
\newcommand\jgr{J.~Geophys.~Res.}
\newcommand\jqsrt{JQSRT}
\newcommand\memsai{MmSAI}
\newcommand\nphysa{NuPhA}
\newcommand\physrep{PhR}
\newcommand\physscr{PhyS}
\newcommand\planss{Planet.~Space~Sci.}
\newcommand\procspie{Proc.~SPIE}

\newcommand\actaa{AcA}
\newcommand\caa{ChA\&A}
\newcommand\cjaa{ChJA\&A}
\newcommand\jcap{JCAP}
\newcommand\na{NewA}
\newcommand\nar{NewAR}
\newcommand\pasa{PASA}
\newcommand\rmxaa{RMxAA}
\newcommand\maps{M\&PS}
\newcommand\aas{AAS Meeting Abstracts}
\newcommand\dps{AAS/DPS Meeting Abstracts}

\bibliographystyle{sciencemag}

%
%
%
%
%
%


\section*{Acknowledgments}
The authors are grateful to Julien Salmon{\color{black}, Robin Canup,} and William Bottke for useful discussions. {\color{black}We gratefully acknowledge the developers of iSALE-3D, including Dirk Elbeshausen, Kai Wünnemann, Gareth Collins and Tom Davison. iSALE-3D models were carried out on the PC cluster and the analysis servers at the Center for Computational Astrophysics, National Astronomical Observatory of Japan.}
\paragraph*{Funding:}
This work was supported by grant 80NSSC20K0422 from the NASA, USA Emerging Worlds program. We also acknowledge funding to the Center for Lunar Origin and Evolution (CLOE), a team in NASA’s SSERVI program (cooperative agreement 80NSSC23M0176)

\paragraph*{Author contributions:}
 H.L$.$ was responsible for Conceptualization, Investigation, Supervision, Project administration, Methodology, Data curation, Validation, Formal analysis, Software, and Visualization. Writing --- original draft, and review and editing. R.D$.$ was responsible for Conceptualization, Investigation, Methodology, Data curation, Validation, Formal analysis, Software, Visualization. Writing --- original draft, and review and editing. K.W$.$ was responsible for Conceptualization, Investigation, Methodology, Funding Acquisition, Data curation, Validation, Supervision, Formal analysis, Software and Project administration.  Writing --- original draft, and review and editing. B.J$.$ was responsible for Conceptualization, Methodology, Resources, Funding Acquisition, Supervision, and Software.  Writing --- original draft, and review and editing. H.C$.$ was responsible for Conceptualization, Investigation, Methodology, Resources, Validation, Formal Analysis, and Visualization.  Writing --- original draft, and review and editing. S.W$.$ was responsible for Investigation, Validation, Formal Analysis, and Software. Writing --- review and editing. R.G.~ was responsible for Formal Analysis.   Writing --- review and editing.

\begin{figure}
\centering
\includegraphics[width=.8\linewidth]{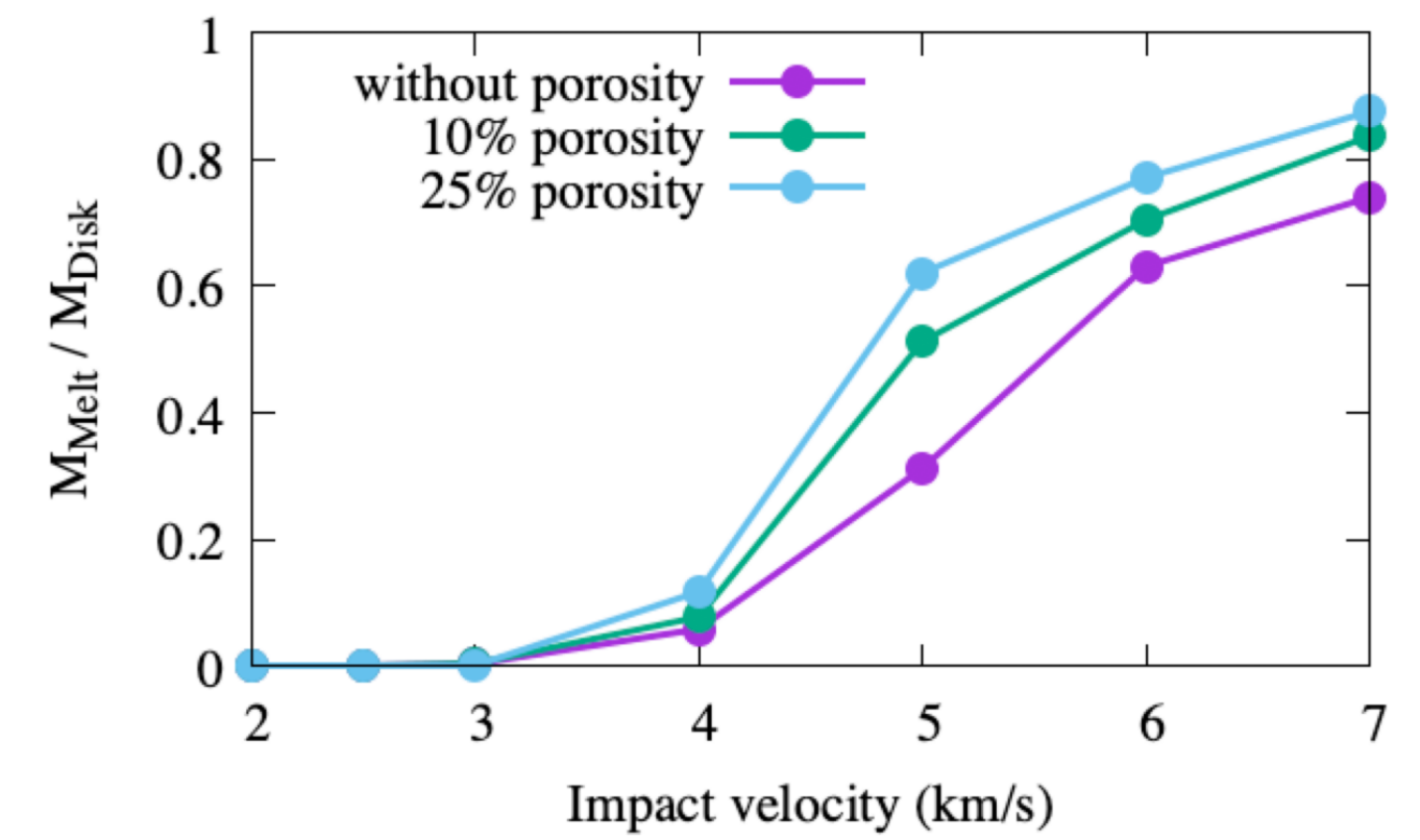}
\caption{\textbf{Melt mass fraction of material ejected with speeds below escape speed for collisions with impacts angles of 45$^\circ$, as a function of impact velocity and porosity.} These simulations are based on those presented in Ref.~\cite{Johnson+2015} and \cite{Wakita+2021}.  Material in this velocity range will evolve into a circum-embryo disk. The melted material cools to become chondrules before they are deposited in the disk. }
\label{fig:Meltfraction}
\end{figure}

\begin{figure*}
\centering
\includegraphics[width=.8\linewidth]{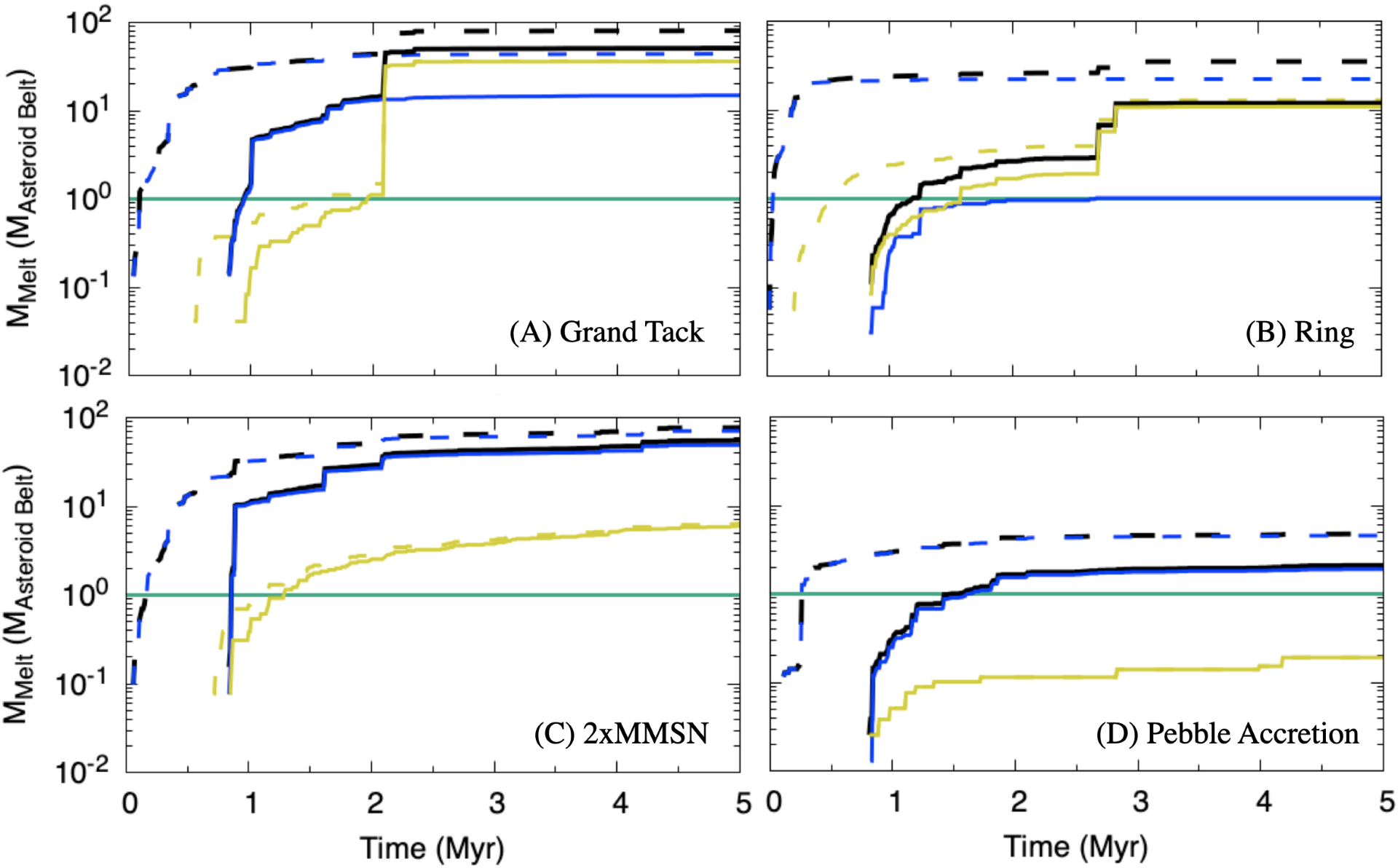}
\caption{\textbf{Cumulative mass in circum-embryos disks formed during embryo-embryo collisions.} The mass is given in terms of the current main asteroid belt mass, $5 \times 10^{-4} M_\oplus$ (hereafter $1 M_{\rm AB}$, the green horizontal lines; \cite{DeMeo&Carry2013,DeMeo&Carry2014,Deienno+2024}.  Panels are from simulations of terrestrial planet accretion based on the following evolutions: (A) Grand Tack as in Ref.~\cite{Walsh+2011}, (B) from the initial conditions as proposed  in Ref.~\cite{Izidoro+2022}(i.e.~Ring), (C) 2xMMSN configuration from \cite{Walsh&Levison2019}, and (D) pebble accretion \cite{Levison+2015}.  The dashed curves show all impacts before 5 Myr, while the solid curves show impacts that occurred after 0.8~Myr.  Black: total amount of mass put into circum-embryo disks. Yellow: Mass in disks containing chondrules with more than 60\% in volume. Blue: Mass in disks containing chondrules with less than 60\% volume.
}
\label{fig:MassTime}
\end{figure*}

\begin{figure}
\centering
\includegraphics[width=.9\linewidth]{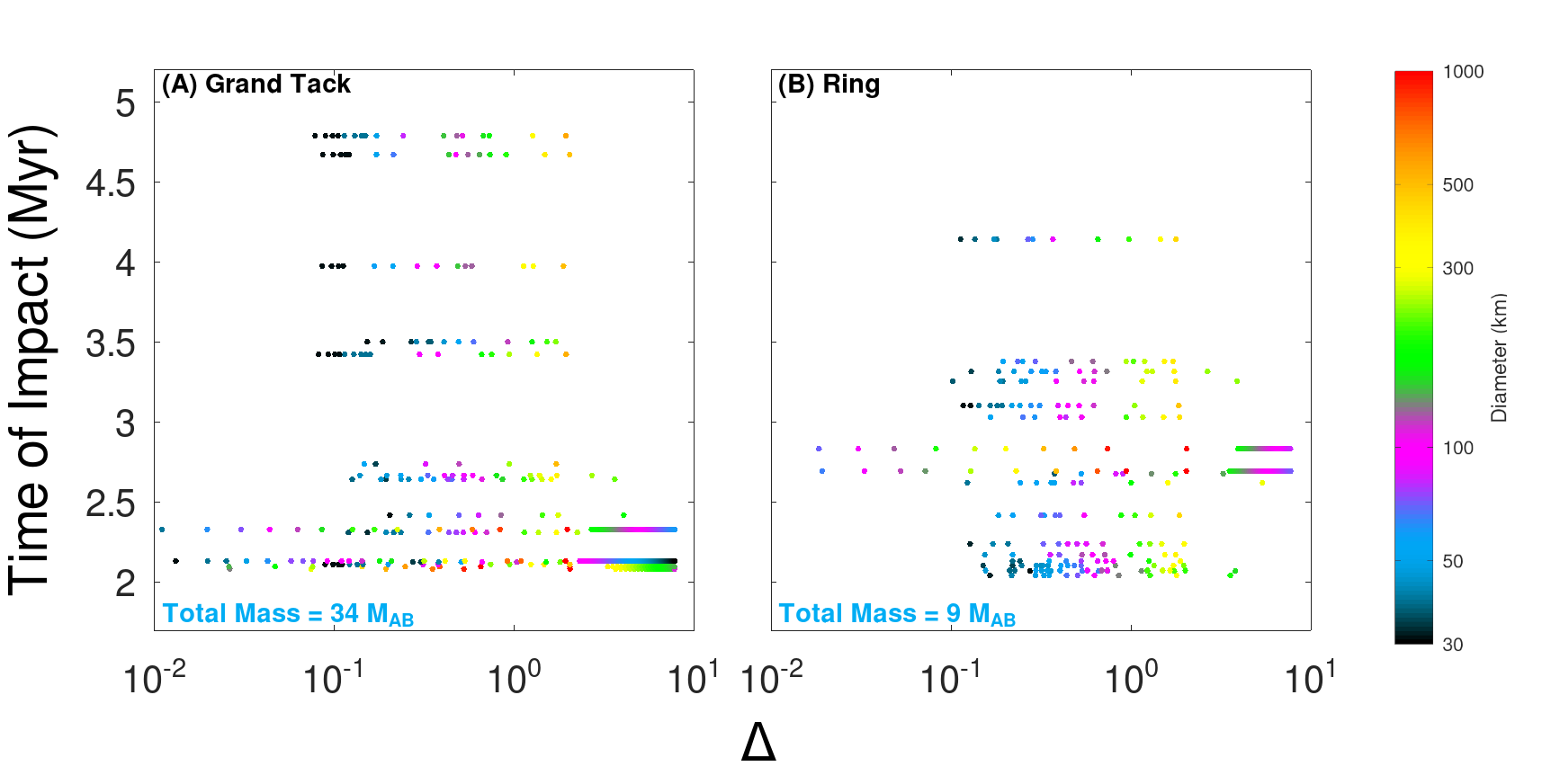}
\caption{\textbf{Size and number of satellites formed in our circum-embryo disks that had more than 60\% melt fraction according to our satellite formation model.}   Panels A and B refer to the satellites that formed during our Grand Tack simulation (Figure~\ref{fig:MassTime}A) and Ring (Figure~\ref{fig:MassTime}B) simulations, respectively. Each horizontal line of points is the product of one disk, where the time of the impact that led to the disk’s creation is indicated on the vertical axis. The bottom horizontal axis is the normalized distance ($\Delta$) from the Roche limit. Only objects with diameters larger than 30 km are plotted, with their size being indicated on the color bar.}
\label{fig:CC12}
\end{figure}

\begin{figure}
\centering
\includegraphics[width=.6\linewidth]{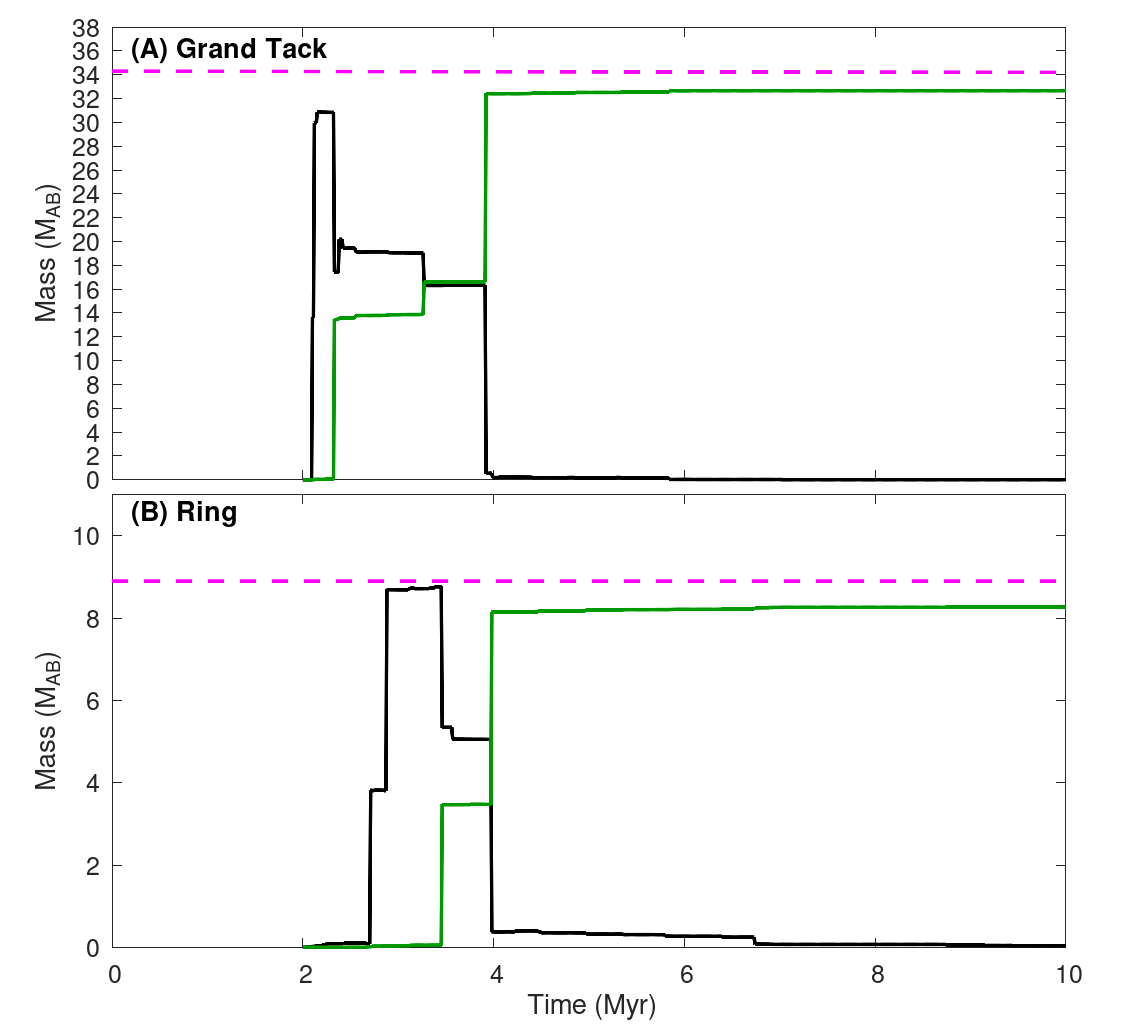}
\caption{\textbf{The temporal evolution of the total mass in chondritic objects in our two terrestrial planet simulations.} A) Grand Tack, and B) Ring.  The black curves show the mass of chondritic satellites, while the green curves show the mass of liberated objects (i.e$.$ those in heliocentric orbit).  The dotted purple curve is the total mass of chondritic satellites produced in the last section and shown in Figure~\ref{fig:CC12}.}
\label{fig:Strip}
\end{figure}

\paragraph*{Competing interests:}
There are no competing interests to declare.
\paragraph*{{\color{black}Data, Code and Materials Availability:}}
{\color{black}All data needed to evaluate the conclusions in this paper are present in the paper or are available in the Dryad
repository at https://doi.org/10.5061/dryad.rfj6q57r7.  In particular,  the data for all six figures is archived in a standard machine-readable format.  In addition, this archive contains that setup files for the new iSALE simulations used in the melt fraction calculations (i.e. Fig~\ref{fig:Meltfraction}), the source codes, scripts, and setup files for the disk mass (Fig~\ref{fig:MassTime}), satellite formation (Fig~\ref{fig:CC12}), and satellite stripping (Fig~\ref{fig:Strip}) calculations.  All setup files are in a standard machine-readable format.  This study did not generate new materials.}

\end{document}